\documentclass[aps,pra,reprint,noeprint,floatfix]{revtex4-1}
\usepackage{bm, braket,amsmath,mathtools,comment,color}
\usepackage{graphicx}
\usepackage[pdfencoding=auto]{hyperref}

\begin{document}
\title{Measurement-induced transitions of the entanglement scaling law in ultracold gases with controllable dissipation}
\author{Shimpei Goto}
\email[]{goto@phys.kindai.ac.jp}
\author{Ippei Danshita}
\email[]{danshita@phys.kindai.ac.jp}
\affiliation{Department of Physics, Kindai University, Higashi-Osaka city, Osaka 577--8502, Japan}
\date{\today}
\begin{abstract}
Recent studies of quantum circuit models have theoretically shown that frequent measurements induce a transition in a quantum many-body system, which is characterized by the change of the scaling law of the entanglement entropy from a volume law to an area law.
In order to propose a way for experimentally observing this measurement-induced transition, we present numerical analyses using matrix-product states on quench dynamics of a dissipative Bose-Hubbard model with controllable two-body losses, which has been realized in recent experiments with ultracold atoms.
We find that when the strength of dissipation increases, there occurs a measurement-induced transition from volume-law scaling to area-law scaling with a logarithmic correction in a region of relatively small dissipation.
We also find that the strong dissipation leads to a revival of the volume-law scaling due to a continuous quantum Zeno effect.
We show that dynamics starting with the area-law states exhibits the strong suppression of particle transport stemming from the ergodicity breaking, which can be used in experiments for distinguishing them from the volume-law states.
\end{abstract}
\maketitle

\section{Introduction}
In a generic quantum many-body system, a pure state is thermalized via long-time evolution, i.e., its expectation values of local observables are very close to those given by a statistical (micro-) canonical ensemble~\cite{rigol_thermalization_2008,nandkishore_many_2015,abanin_colloquium_2019}.
The entanglement entropy of such a thermal pure state obeys volume-law scaling, corresponding to the fact that the entropy of a thermal density matrix is extensive~\cite{eisert_textitcolloquium_2010,nakagawa_universality_2018,garrison_does_2018}.
Recent advances in understanding and controlling coherent quantum many-body dynamics have revealed a few exceptional systems which do not show such thermalization.
First, in integrable systems, such as the Lieb-Liniger model and the one-dimensional (1D) Ising model with a transverse field, many integrals of motion prevent a pure state from relaxation towards a thermal state~\cite{kinoshita_quantum_2006,rigol_relaxation_2007,rigol_thermalization_2008,vidmar_generalized_2016}.
Second, in many-body localized (MBL) systems, disordered potentials forbid ballistic propagation of quantum information such that the entanglement entropy grows only logarithmically with time~\cite{znidaric_many-body_2008,bardarson_unbounded_2012,nandkishore_many_2015,abanin_colloquium_2019}.

Recent theoretical studies of quantum circuit models have proposed another class of exceptional systems~\cite{li_quantum_2018,chan_unitary-projective_2019,skinner_measurement-induced_2019,szyniszewski_entanglement_2019,jian_measurement-induced_2020,gullans_dynamical_2019,li_measurement-driven_2019,choi_quantum_2020,bao_theory_2020,tang_measurement-induced_2020}.
In these studies, random unitary dynamics with probabilistic measurements have been investigated.
It has been shown that when the probability of measurements increases, the scaling law of the entanglement entropy exhibits a transition from a volume law to an area law at a certain critical point.
Since the volume-law scaling is a necessary condition for a pure state to be thermal, the emergence of the area-law scaling means that many measurements prevent a state after long-time evolution from the thermalization.
Despite the intensive interest in this measurement-induced transition (MIT), its experimental observation is still lacking.
In order to observe the MIT, one needs an experimental system with long coherence time and high controllability of measurements.

Ultracold gases have served as an ideal platform for analyzing long-time coherent dynamics of many-body systems thanks to their long thermalization time and isolation from the environment.
Indeed, coherent quantum dynamics of integrable systems~\cite{kinoshita_quantum_2006} and MBL systems~\cite{schreiber_observation_2015} has been observed in this platform for the first time.
Recent experiments have successfully introduced controllable dissipation to ultracold-gas systems to create and manipulate quantum many-body states~\cite{barontini_controlling_2013,patil_measurement-induced_2015,luschen_signatures_2017,tomita_observation_2017,bouganne_anomalous_2020}.
Since the introduced dissipation corresponds to a continuous quantum measurement, which can be interpreted as probabilistic measurements in terms of quantum trajectory representation of open quantum systems, we expect that it may be utilized for causing the MIT.

In this paper, we propose a specific protocol to realize the MIT with use of ultracold gases in optical lattices.
By means of the quantum trajectory method implemented with matrix product states (MPS)~\cite{dum_monte_1992,daley_quantum_2014,schollwock_density-matrix_2011}, we analyze the 1D Bose-Hubbard model with two-body losses, which can be widely controlled in experiment by the strength of a photoassociation (PA) laser~\cite{tomita_observation_2017}.
We find that this system exhibits a MIT from volume-law scaling to area-law scaling with a logarithmic correction (ALSLC) when the strength of the two-body losses increases in a weakly dissipative regime.
Moreover, we find another MIT in a strongly dissipative regime.
The latter transition can be attributed to a continuous quantum Zeno effect (QZE) and has not been reported in previous literature studying quantum circuit models.
We show that the experimentally accessible momentum distribution reflects the changes of the scaling laws.
We also analyze dynamics after release of the particles to an empty space in order to show that the states with ALSLC can be distinguished from the volume-law states by observing the strong suppression of particle transport, which can be recognized as the tendency towards the ergodicity breaking.

The rest of the paper is organized as follows.
In Sec.~\ref{sec:method}, we define the master equation describing ultracold bosons with a PA laser in a 1D optical lattice, and introduce the quantum trajectory method for analyzing the master equation. We also define ``entanglement entropy'' used in this study in the section. In Sec.~\ref{sec:transition}, we show that there exist two MIT in this system and that the MIT has the reentrant structure. In Sec.~\ref{sec:exp}, we discuss how to detect the MIT in ultracold-gas experiments. In Sec.~\ref{sec:summary}, we summarize the results.

\section{Model and methods\label{sec:method}}
We consider ultracold bosons confined in an optical lattice.
We assume that the lattice potential in the transverse (yz) directions is so deep that the hopping in these direction is forbidden, i.e., the system is 1D. 
We also assume that the lattice potential in the longitudinal (x) direction is deep enough for the tight-binding approximation to be valid.
The two-body losses can be introduced by exposing the system to a PA laser~\cite{tomita_observation_2017}, which couples a local two-atom state to a molecular state with a very short lifetime.
In this system, the time-evolution of a density matrix \(\hat{\rho}(t)\) can be effectively described by the master equation in Lindblad form~\cite{tomita_observation_2017,gorini_completely_1976,lindblad_generators_1976}
\begin{align}
    \frac{\mathrm{d}}{\mathrm{d} t} \hat{\rho}(t) &= -\frac{\mathrm{i}}{\hbar}[\hat{H}, \hat{\rho}(t)] + \hat{L}[\hat{\rho}(t)]
    \label{eq:master_eq}
\end{align}
with the 1D Bose-Hubbard Hamiltonian
\begin{align}
    \label{eq:BoseHubbard}
    \hat{H} = -J\sum^{M-1}_{i=1} (\hat{b}^\dagger_i \hat{b}_{i+1} + \mathrm{H.c.}) + \frac{U}{2} \sum^M_{i=1} \hat{n}_i(\hat{n}_i - 1),
\end{align}
and the Lindblad superoperator for two-body atom losses
\begin{align}
    \label{eq:Lindblad}
    \hat{L}[\hat{\rho}] = - \frac{\gamma}{2} \sum_i (\hat{b}^\dagger_i \hat{b}^\dagger_i \hat{b}_i \hat{b}_i \hat{\rho} + \hat{\rho} \hat{b}^\dagger_i \hat{b}^\dagger_i \hat{b}_i \hat{b}_i - 2\hat{b}_i\hat{b}_i\hat{\rho}\hat{b}^\dagger_i\hat{b}^\dagger_i).
\end{align}
Here, \(J\) is the hopping amplitude, \(M\) is the number of lattice sites, \(\hat{b}^\dagger_i\) (\(\hat{b}_i\)) creates (annihilates) a boson at site \(i\), \(U\) is the on-site Hubbard interaction, \(\hat{n}_i = \hat{b}^\dagger_i \hat{b}_i\), and \(\gamma \) is the strength of the two-body inelastic collision which can be controlled by the intensity of the PA laser.
We denote the number of remaining particles in the system as \(N\), i.e., \(N = \sum_i \braket{\hat{n}_i}\).
At initial time \(t=0\), we assume that the system is a Mott insulating state at unit filling, i.e., \(\ket{\psi_0} = \prod_i \hat{b}^\dagger_i\ket{0}\) and thus \(\hat{\rho}(0) = \ket{\psi_0}\bra{\psi_0}\), where \(\ket{0}\) denotes the vacuum state.

Solving the master equation~\eqref{eq:master_eq} requires a very high numerical cost in general because the number of coefficients in the density matrix is the square of the dimension of the Hilbert space. 
To circumvent this difficulty, we use quantum trajectory techniques which treat pure states in the density matrix~\cite{daley_quantum_2014,dum_monte_1992} instead of treating the density matrix directly.
Following the quantum trajectory techniques, we calculate the time-evolved state 
\begin{align}
    \ket{\psi(t)} = \mathrm{e}^{-\mathrm{i}\frac{t}{\hbar}\hat{H}_\mathrm{eff}}\ket{\psi_0}
\end{align}
with the effective non-hermitian Hamiltonian 
\begin{align}
    \hat{H}_\mathrm{eff} = \hat{H} - \mathrm{i}\frac{\hbar\gamma}{2} \sum_i \hat{b}^\dagger_i \hat{b}^\dagger_i \hat{b}_i \hat{b}_i.
\end{align}
As time \(t\) increases, the norm of the time-evolved state \(\ket{\psi(t)}\) decreases because of the non-hermitian part of the effective Hamiltonian \(\hat{H}_\mathrm{nh}\).
When the squared norm of the time-evolved state becomes lower than a random number generated from the uniform distribution \((0, 1)\), we calculate a probability \(p_i \propto \braket{\psi(t)|\hat{b}^\dagger_i\hat{b}^\dagger_i\hat{b}_i\hat{b}_i|\psi(t)}\) and choose one index \(j\) according to the probability \(p_i\).
Then, the jump operator \(\hat{b}_j \hat{b}_j\) is applied to \(\ket{\psi(t)}\) and the state is normalized.
This stochastic process emulates the open dynamics described by the master equation in Lindblad form, and the expectation values are obtained by the sample average
\begin{align}
    \begin{aligned}
    \braket{\hat{O}(t)} &= \mathrm{Tr}[\hat{O}\hat{\rho}(t)] \\
                        &\simeq \frac{1}{K}\sum^K_{l=1}\frac{\braket{\psi_l(t)|\hat{O}|\psi_l(t)}}{\braket{\psi_l(t)|\psi_l(t)}},
    \end{aligned}
\end{align}
where \(\ket{\psi_l(t)}\) is the \(l\)-th sample of the stochastic process and \(K\) is the number of samples.
Notice that the application of the jump operator and the subsequent normalization correspond to a quantum measurement.
In the sense that a series of the measurement events stemming from the dissipation occur probabilistically according to the random number and the probability distribution \(p_i\), the dissipation can be interpreted as probabilistic measurements.

For numerically efficient calculations in 1D, we represent a state \(\ket{\psi(t)}\) with MPS and perform the time evolution by means of the time-evolving block decimation algorithm~\cite{vidal_efficient_2003,vidal_efficient_2004,white_real-time_2004,daley_time-dependent_2004} using the optimized Forest-Ruth-like decomposition~\cite{omelyan_optimized_2002}.
The truncation error is set to be less than \(10^{-8}\), and the time step \(\Delta t\) is adaptively changed after each jump operation as \(\Delta t = \min \{-\log(0.9)\hbar/\braket{\psi(t)|\mathrm{i}\hat{H}_\mathrm{nh}|\psi(t)}, \Delta t_{\max}\} \) in order to avoid a rapid decrease in the norm of wavefunction.
Here, \(\Delta t_{\max}\) is the upper bound of the time step that we set to be \(0.05 \hbar/J\) (\(0.02 \hbar/J\)) for small to intermediate \(\hbar \gamma /J\) (large \(\hbar \gamma / J \geq 100 \)).

It should be cautioned that we have to define what we call ``entanglement entropy'' in this study because the ordinary entanglement entropy is defined only for pure states \(\ket{\phi}\) on a system biparted into subsystems \(A\) and \(B\) as 
\begin{align}
    S_A(t) = -\mathrm{Tr}\hat{\rho}_A(t) \ln \hat{\rho}_A(t),
\end{align}
where \(\hat{\rho}_A\) is a reduced density matrix defined as 
\begin{align}
    \hat{\rho}_A(t) = -\mathrm{Tr}_B\ket{\phi(t)}\bra{\phi(t)}.
\end{align}
Here, \(\mathrm{Tr}_B\) means a partial trace over the subsystem \(B\).
In this study, as well as other studies investigating the MIT, the statistical average of the entanglement entropy of \(\ket{\psi(t)}/\sqrt{\braket{\psi(t)|\psi(t)}}\) is called ``entanglement entropy'' and the size dependence of the ``entanglement entropy'' is discussed.
In other words, what we discuss is typical behaviors of the entanglement entropy of relevant states in a density matrix \(\hat{\rho}(t)\).
An equal bipartition does not always give the maximal entanglement entropy in the presence of the two-body loss. 
Therefore, we define the average of the maximal bipartite entanglement entropy
\begin{align}
    S_{\max}(t) = \Braket{\max_A S_A(t)},
\end{align}
where \(\max_A\) means the biparted subsystem \(A\) that gives the maximal entanglement entropy.
In this study, we discuss the scaling law of the ``entanglement entropy'' based on \(S_{\max}(t)\).
Hereafter, we call \(S_{\max}(t)\) as entanglement entropy for simplicity.

It is worth noting that the MIT in the Bose-Hubbard model~\eqref{eq:BoseHubbard} with local projective measurements has been studied in Ref.~\cite{tang_measurement-induced_2020}.
In contrast to the previous study, here we incorporate the specific form of controllable dissipation that has been experimentally realized and show an observable suited for characterizing the transitions.

\section{Measurement-induced transitions in the dissipative Bose-Hubbard models\label{sec:transition}}
\begin{figure}[htp]
    \centering
    \includegraphics[width=\linewidth]{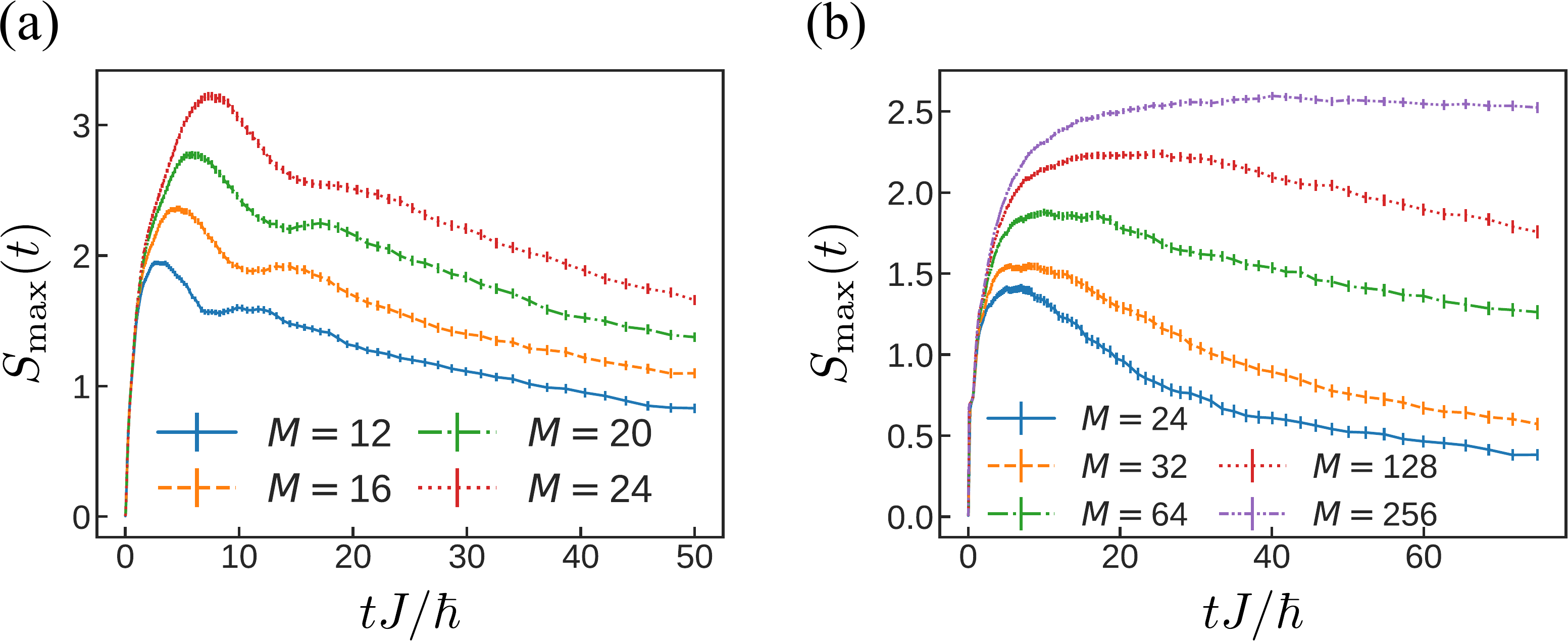}
    \caption{Time evolution of the entanglement entropy for (a)\(\hbar \gamma / J =0.5\) and (b) \(\hbar \gamma / J = 5.0\) for several system sizes at \(U/J = 5.0\). Error bars indicate \(1\sigma \) uncertainty.~\label{fig:time-evol}}
\end{figure}

Figure~\ref{fig:time-evol} shows the time-evolution of the entanglement entropy for different values of \(\hbar \gamma / J\) and \(M\) at \(U/J = 5.0\).
By comparing the case of \((\hbar \gamma / J, M) = (0.5, 24)\) with that of \((\hbar \gamma / J, M) = (5.0, 24)\), we see that the dissipation suppresses the growth of the entanglement entropy.
Thanks to this suppression, when \(\hbar \gamma / J = 5.0\), we can compute long-time dynamics of a relatively large system, say \(M=256\).
The general tendency of the entanglement entropy in the presence of the two-body losses is that it rapidly grows in a short time regime and gradually decreases due to the two-body losses after taking a maximal value.
We show below that the maximal entanglement entropy during the time evolution at \(\hbar \gamma / J = 5.0\) obeys ALSLC.
In Fig.~\ref{fig:time-evol}(b), we see that a steady-value region, where \(S_{\max}(t)\) takes almost the same value as the maximal value, develops when the system size increases (see, e.g., the region \(15 \lesssim tJ/\hbar \lesssim 30\) in the case of \((\hbar \gamma / J, M) = (5.0, 128)\)).
The presence of the steady-value region allows us to identify the states with ALSLC analyzed in the present work as those in the realm of the MIT~\cite{li_quantum_2018,chan_unitary-projective_2019,skinner_measurement-induced_2019}.

\begin{figure}[htp]
    \centering
    \includegraphics[width=\linewidth]{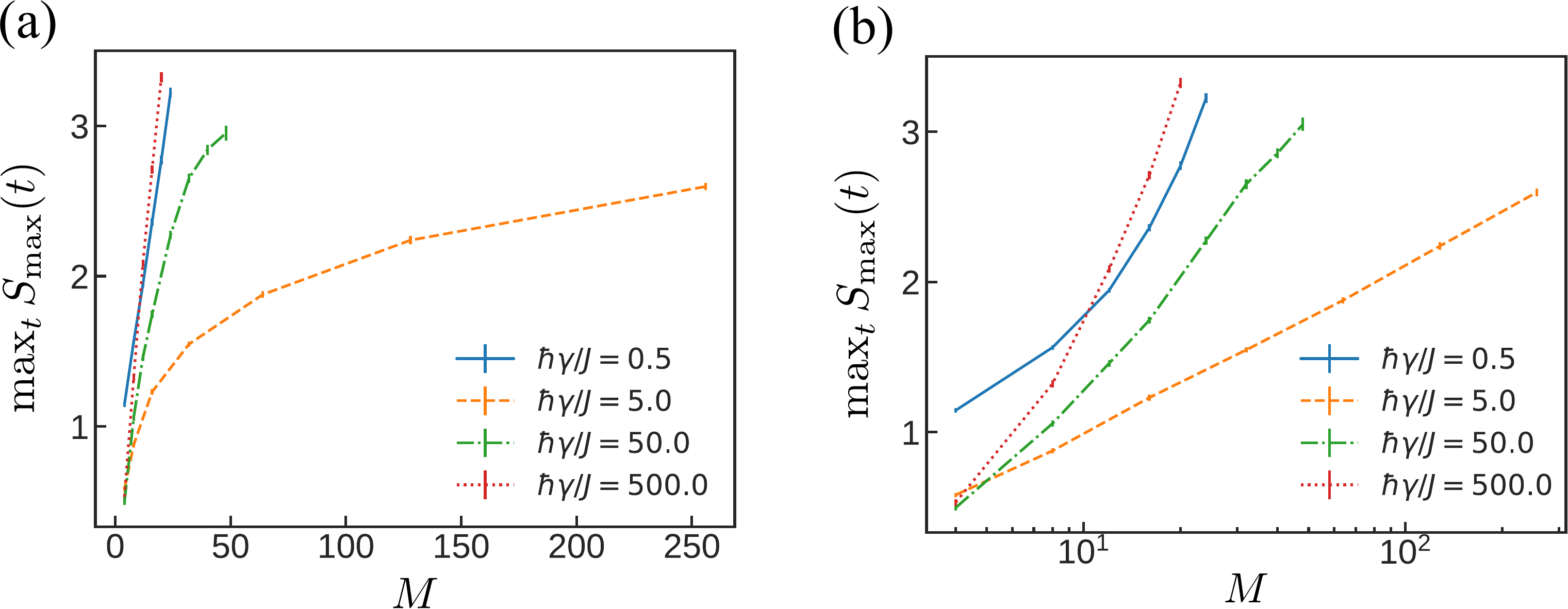}
    \caption{System size \(M\) dependence of the maximal entanglement entropy \(\max_t S_{\max}(t)\) (a) in the linear scale for \(M\) and (b) in the logarithmic scale for \(M\) at \(U/J = 5.0\). The blue solid, orange dashed, green dashed-dotted, and red dotted lines correspond to \(\hbar \gamma / J = 0.5\), \(5.0\), \(50.0\), and \(500.0\), respectively. Error bars indicate \(1\sigma \) uncertainty.~\label{fig:Size_dep}}
\end{figure}

Figure~\ref{fig:Size_dep} shows the maximal values of the entanglement entropy, \(\max_t S_{\max}(t)\), during the time evolution as a function of the system size \(M\) for \(\hbar \gamma / J = 0.5\), \(5.0\), \(50.0\), and \(500.0\) at \(U/J = 5.0\). 
When the dissipation is as small as \(\hbar \gamma / J = 0.5\) or is as large as \(\hbar \gamma / J = 500.0\), the entanglement entropy grows linearly with \(M\) within the system size that we can numerically compute (\(M \simeq 24\)), i.e., it follows the volume-law scaling.
On the contrary, in an intermediate dissipation regime, including \(\hbar \gamma / J = 5.0\) and \(50.0\), the entanglement entropy grows logarithmically with \(M\), i.e., it follows ALSLC.
We call the scaling with the logarithmic correction as area law in the sense that the correction grows more slowly with the system size than algebraic growth, i.e., it is not extensive.
This observation means that when the strength of the dissipation increases, the system exhibits a transition from a volume-law state to an ALSLC state at a relatively small value of dissipation and the other transition to another volume-law state at a relatively large value.
In short, in the present system the MIT has the reentrant structure.

The presence of the volume-law state at the small dissipation, \(\hbar \gamma / J =0.5\), implies that there is a finite critical value \({(\hbar \gamma/J)}_\mathrm{c}\) for the MIT likewise the cases of random unitary circuits.
On the other hand, that at the large dissipation, \(\hbar \gamma / J = 500.0\), can be interpreted as a consequence of the continuous QZE.\@
More specifically, the strong two-body losses suppress double occupation at each site such that the particles in the system behave as hardcore bosons~\cite{garcia-ripoll_dissipation-induced_2009}.
Hence, after several loss events at an early time range, which create a considerable number of holes, the measurement events rarely happen so that the holes spread ballistically to lead to the volume-law entanglement.

\begin{figure}
    \centering
    \includegraphics[width=0.8\linewidth]{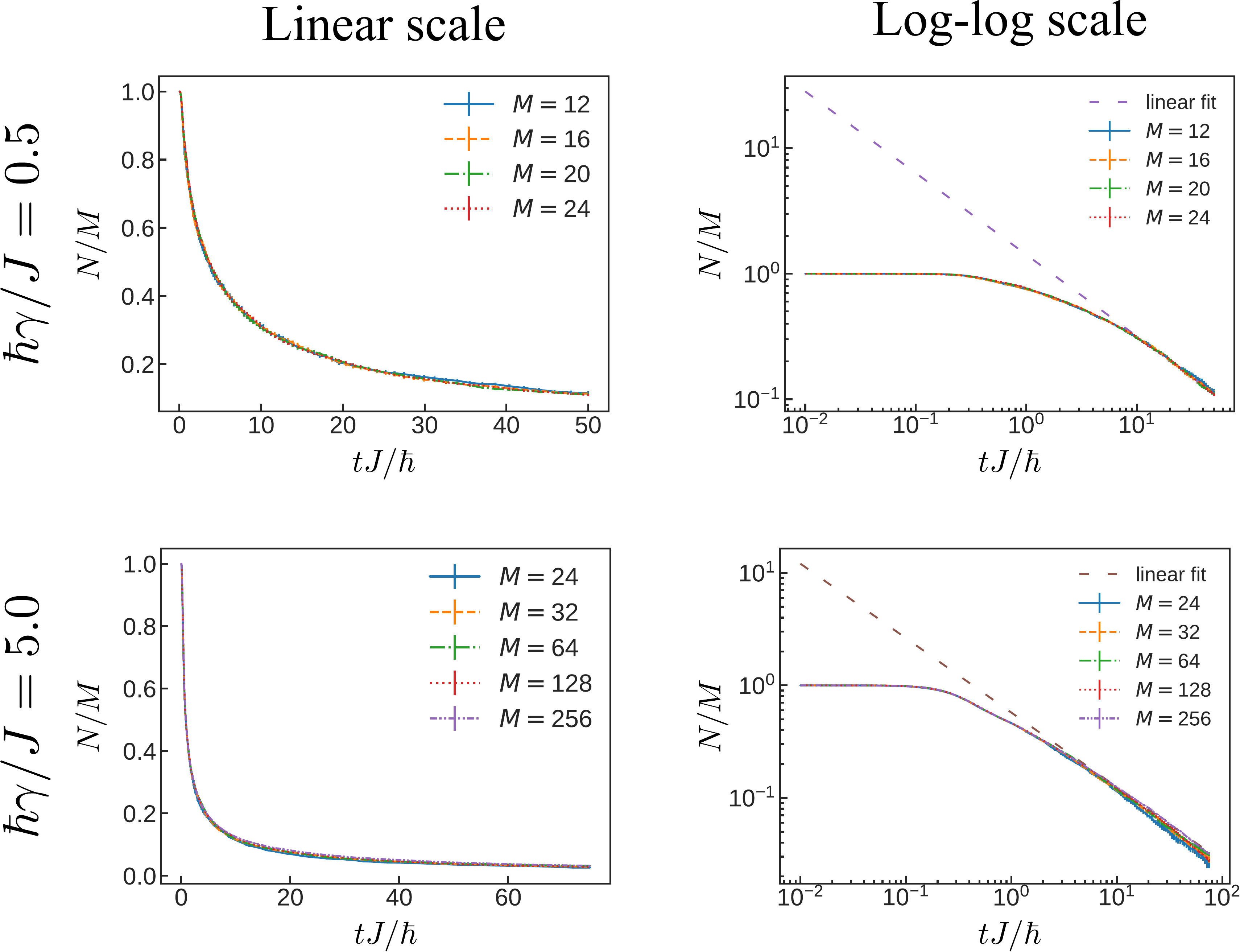}
    \caption{Time evolutions of the number of remaining particles per site for several system sizes with \(\hbar \gamma / J = 0.5\) and \(\hbar \gamma / J = 5.0\). The Hubbard interaction \(U/J\) is set to 5.0. Linear fits are obtained from the data for \(t > 10 \hbar / J\) (\(t > 5.0 \hbar / J\)) in the largest \(M\) for \(\hbar \gamma / J = 0.5\) (\(\hbar \gamma / J = 5.0\)) case. Error bars indicate 1\(\sigma \) uncertainty.~\label{fig:Nt}}
\end{figure}

\begin{figure}
    \centering
    \includegraphics[width=\linewidth]{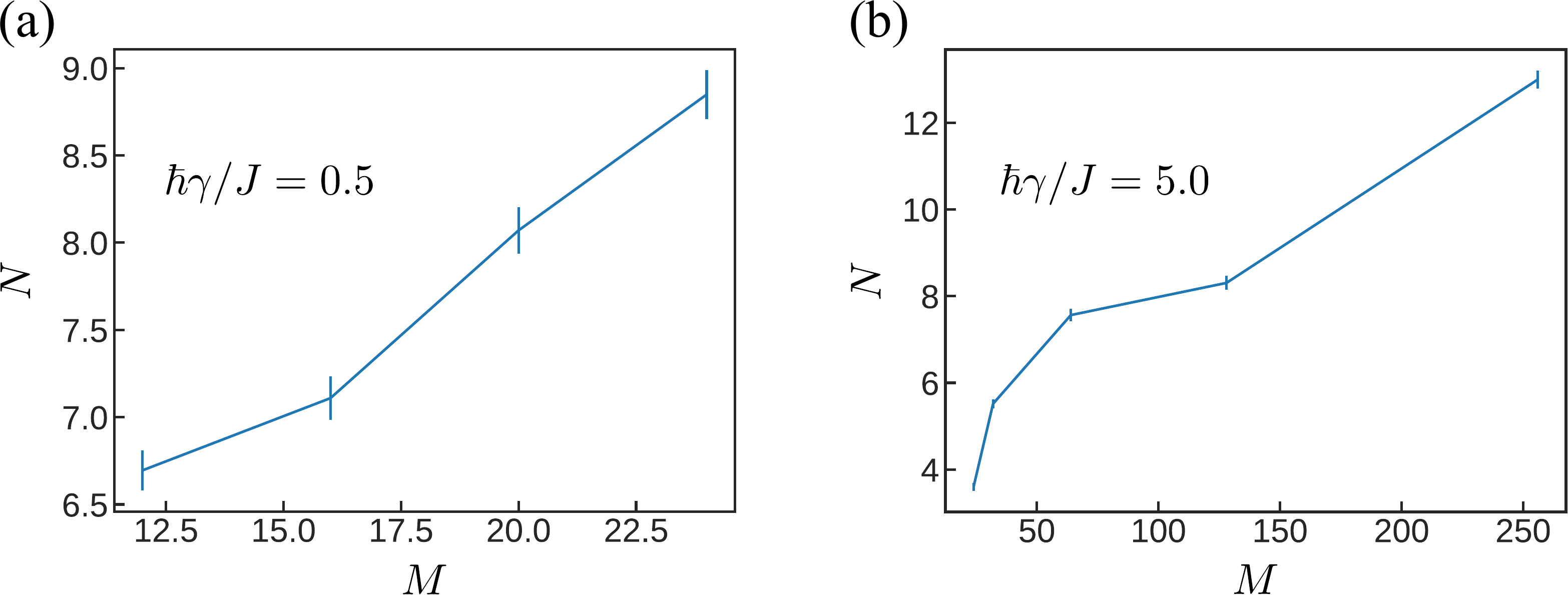}
    \caption{The system size \(M\) dependencies of the number of remaining particles \(N\) when the entanglement entropy takes the maximal value for (a) \(\hbar \gamma / J = 0.5\) and (b) \(\hbar \gamma / J = 5.0\). Error bars indicate 1\(\sigma \) uncertainty.~\label{fig:NvsM}}
\end{figure}

Since the number of remaining particles \(N\) continues to decrease in the Lindblad dynamics, one might suspect that the transitions of the scaling law are results of the decrease of particles.
Figure~\ref{fig:Nt} represents the time evolution of the average density \(N/M\) in the dynamics shown in Fig.~\ref{fig:time-evol}.
For both \(\hbar \gamma / J = 0.5\) and \(5.0\) cases, the density decreases algebraically in long-time dynamics and its dependence on the system size is almost absent.
The exponents of the algebraic decreases estimated from the linear fits are \(-0.65\) for \(\hbar \gamma / J = 0.5\) and \(-0.66\) for \(\hbar \gamma / J = 5.0\).
These exponents are almost the same before and after the scaling law transition.
Furthermore, as shown in Fig.~\ref{fig:NvsM}, the number of remaining particles when the entanglement entropy takes the maximal value increases almost linearly as the system size increases for both \(\hbar \gamma / J = 0.5\) and \(5.0\) cases.
Therefore, the scaling of \(N\) is not so different before and after the transitions of the scaling laws, and thus the transitions cannot be understood as the result of the decrease of particles.

\begin{figure}
    \centering
    \includegraphics[width=\linewidth]{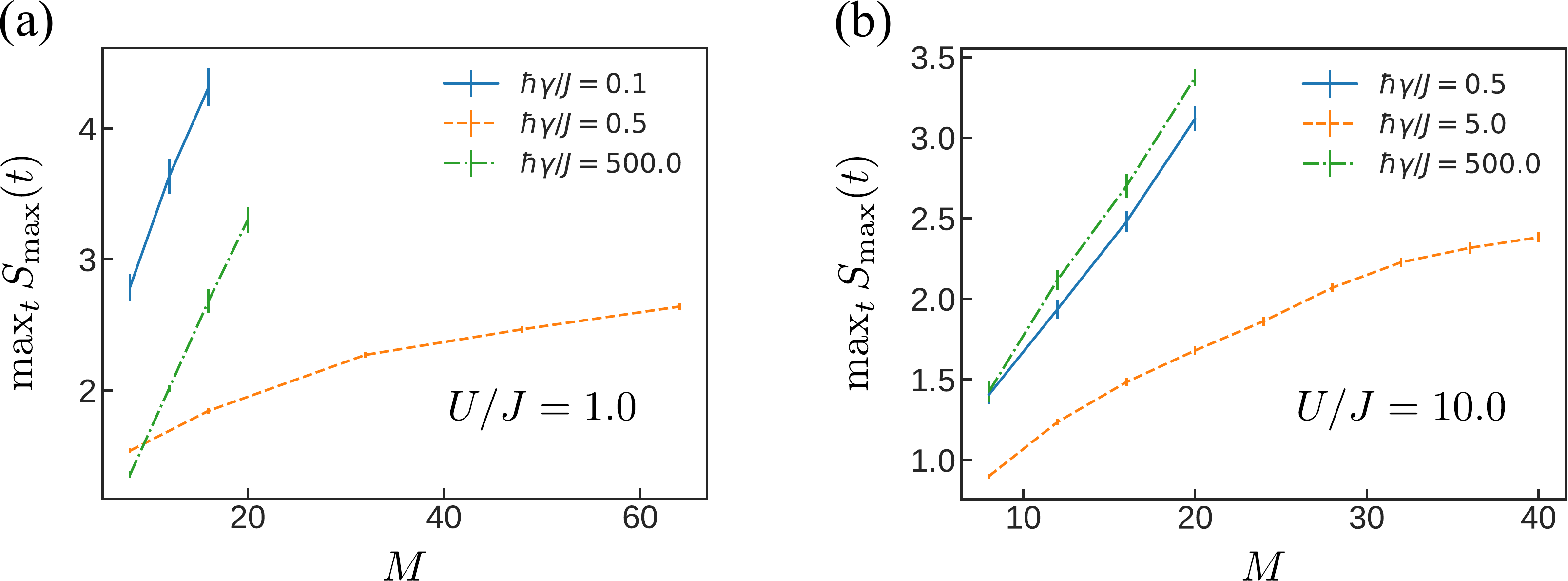}
    \caption{System size \(M\) dependencies of the maximal entanglement entropy \(\max_t S_{\max}(t)\) (a) with \(U/J = 1.0\) and (b) with \(U/J = 10.0\) for several strengths of the dissipation \(\gamma \). Error bars indicate 1\(\sigma \) uncertainty.\label{fig:Udep}}
\end{figure}

The reentrant structure we found is present in a broad range of \(U/J\).
Figure~\ref{fig:Udep} represents the system size dependencies of the maximal value of the entanglement entropy for \(U/J = 1.0\) and \(U/J = 10.0\) cases.
One can see the reentrant structure for both cases.
For the \(U/J = 1.0\) case, even with a small dissipation \(\hbar \gamma / J = 0.5\), the scaling law of the entanglement entropy is ALSLC in contrast to the \(U/J = 5.0\) case.
This can be attributed to the fact that the double occupancy rate increases compared to the \(U/J = 5.0\) case and thus the probability of measurement is effectively increased.

\section{How to experimentally detect measurement-induced transitions\label{sec:exp}}
In closed systems, a kind of the entanglement entropy, namely the 2nd order R\'{e}ny entropy, has been observed in experiments with ultracold gases in optical lattices by preparing a copy of the target system and measuring interference between the target and the copy~\cite{islam_measuring_2015,kaufman_quantum_2016}.
However, in open systems with dissipation, it is hard to use the same protocol because the copy cannot perfectly mimic measurement events which happen in a stochastic manner.
Hence, it is imperative to point out alternative experimental observables that can distinguish the ALSLC states from the volume-law states.

\subsection{Momentum distribution}
In this subsection, we show that the momentum distribution 
\begin{align}
    \braket{\hat{n}_k} = \frac{1}{M}\sum_{ij}\braket{b^\dagger_i b_j}\mathrm{e}^{\mathrm{i} k(i-j)},
\end{align}
which is a standard observable in ultracold-gas experiments, reflects the scaling law of the entanglement entropy.
Here, we set the lattice spacing to unity.
{We set \(U/J\) to \(5.0\) in this subsection.

\begin{figure}
    \centering
    \includegraphics[width=0.5\linewidth]{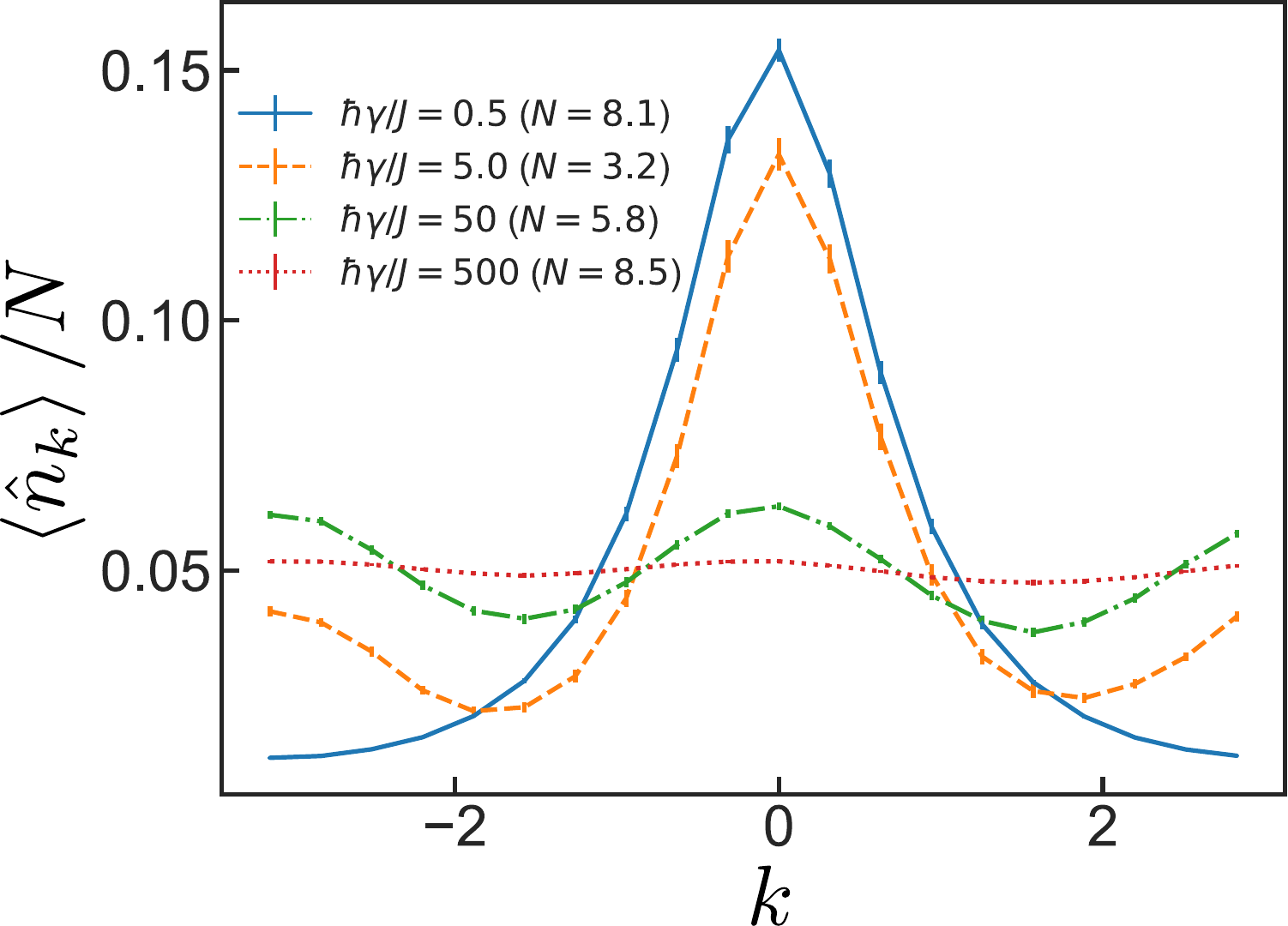}
    \caption{Normalized momentum distributions for several values of the dissipation strength \(\gamma \) at the time that gives \(\max_t S_{\max}(t)\). The blue solid, orange dashed, green dashed-dotted, and red dotted lines correspond to \(\hbar \gamma / J = 0.5\), \(5.0\), \(50.0\), and \(500.0\), respectively. The system size \(M\) is set to 20 in order to investigate a vast range of \(\gamma \). Error bars indicate \(1\sigma \) uncertainty.~\label{fig:Momentum}}
\end{figure}

Figure~\ref{fig:Momentum} shows the normalized momentum distributions for \(\hbar \gamma / J = 0.5\), \(5.0\), \(50.0\), and \(500.0\) at the time that gives \(\max_t S_{\max}(t)\) (See Sec.~\ref{sec:method} for the definition).
The system size is set to \(M = 20\) in order to compute states with the volume-law entanglement.
In each of the three different regions of the dissipation strength, \(\braket{\hat{n}_k}/N\) exhibits a distinct signal.
Here, \(N\) is the total number of remaining particles in the system.
In the case of the small dissipation, \(\hbar \gamma / J = 0.5\), there exists a single peak at \(k=0\).
In the intermediate region, including \(\hbar \gamma / J = 5.0\) and \(50.0\), the dips at \(|k| = \pi/2\) are developed.
In the case of the strong dissipation, \(\hbar \gamma / J = 500.0\), the distribution is almost flat.
In order to characterize the signals more quantitatively, we show in Fig.~\ref{fig:Visibility} the visibility \(\braket{\hat{n}_\pi}/\braket{\hat{n}_{\pi/2}}\) as a function of \(\hbar \gamma / J\).
\begin{figure}
    \centering
    \includegraphics[width=0.5\linewidth]{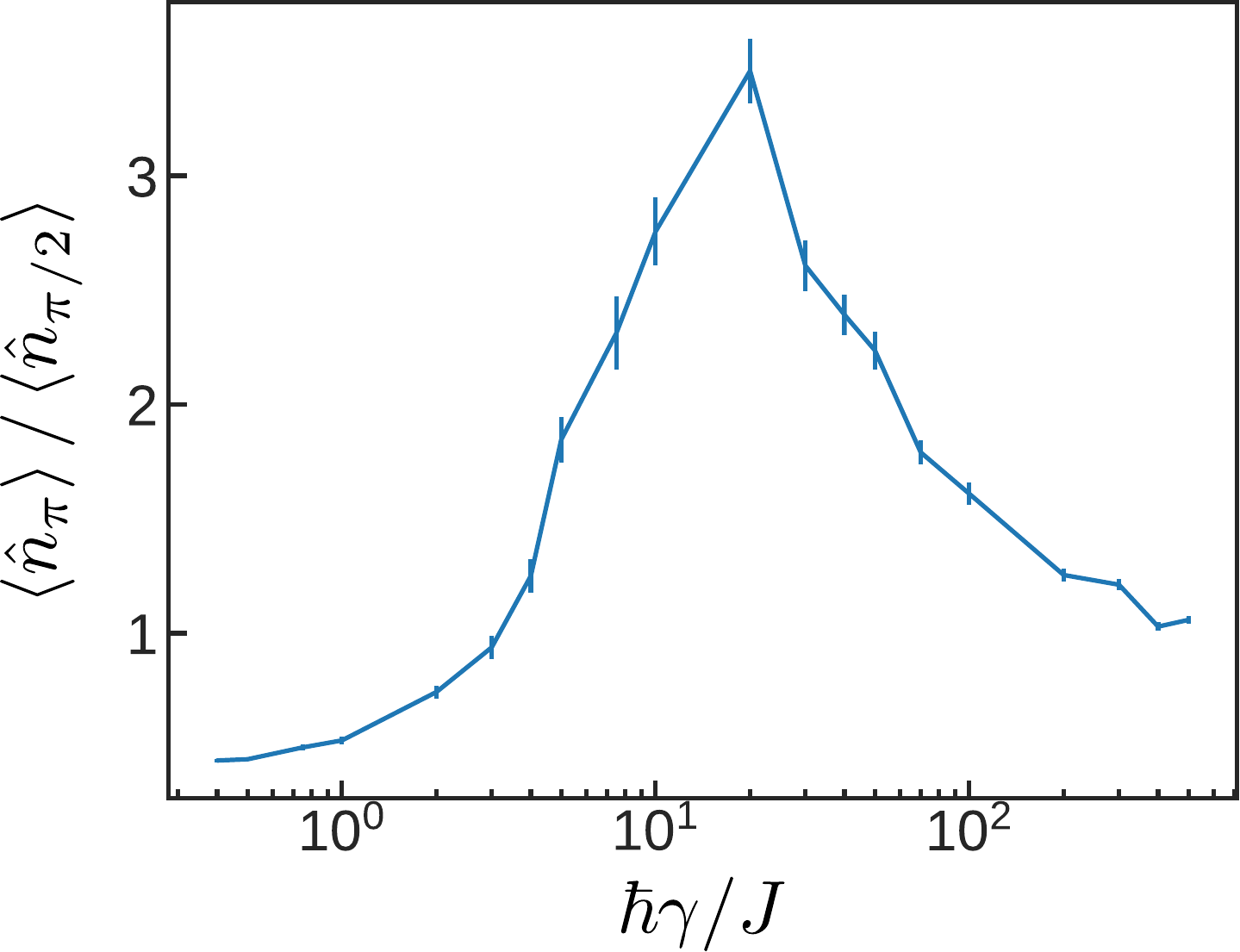}
    \caption{Visibility \(\braket{\hat{n}_\pi} / \braket{\hat{n}_{\pi/2}}\) as a function of the dissipation strength \(\gamma \). Although it is in practice impossible to precisely determine the critical points with our matrix product states method, we have checked that the states in the region \(2 \leq \hbar \gamma / J \leq 50\) safely obey ALSLC. Error bars indicate \(1\sigma \) uncertainty.~\label{fig:Visibility}}
\end{figure}
Since the visibility becomes considerably large in the intermediate region, where the states with ALSLC emerge, it can be used for distinguishing the states with ALSLC from the volume-law states.
Notice that the visibility at \(M=20\) shown in Fig.~\ref{fig:Visibility} does not exhibit any singular behaviors across the transition points because the system size is too small.

\begin{figure}
    \centering
    \includegraphics[width=0.8\linewidth]{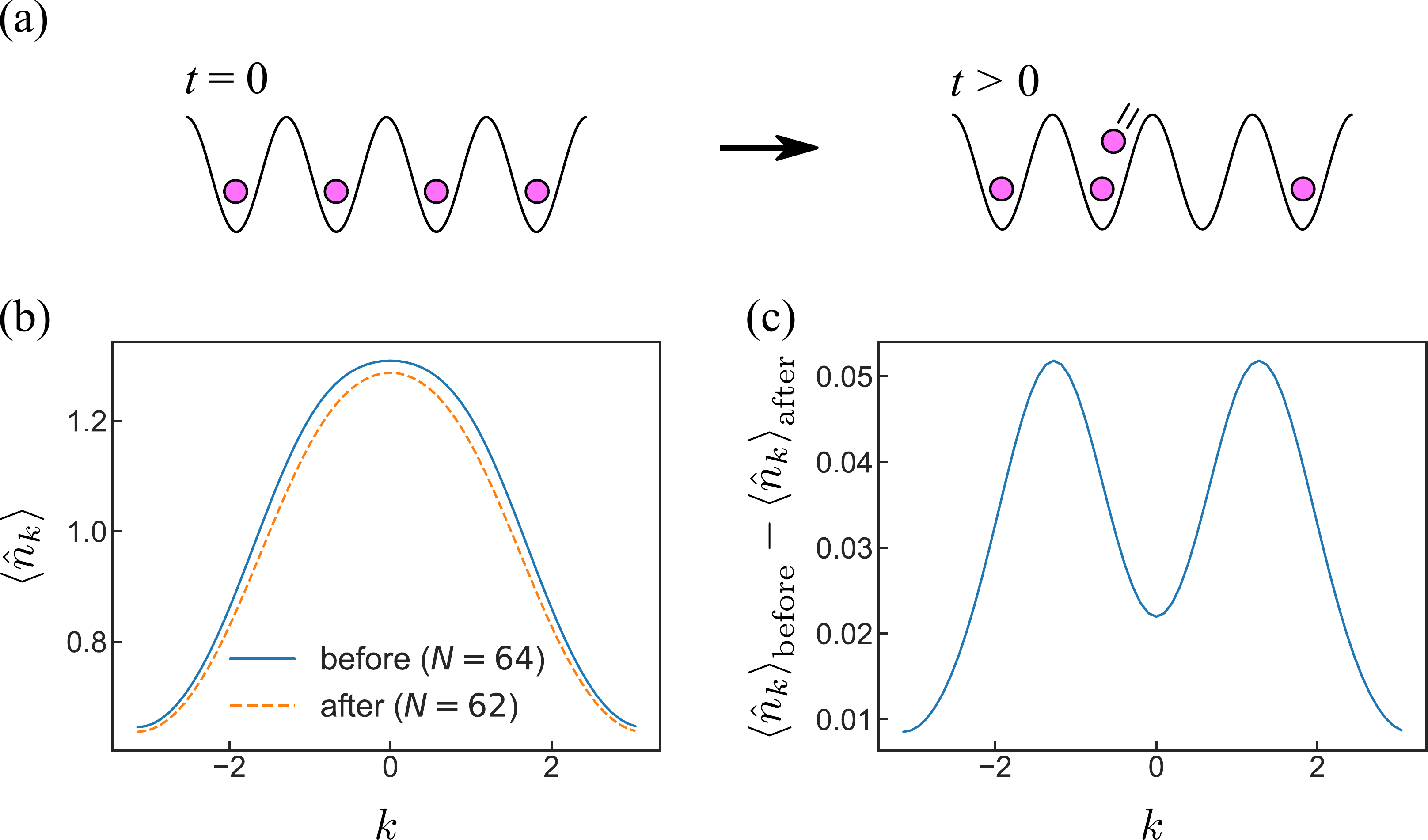}
    \caption{(a) A possible loss process in the early time dynamics. Atoms with finite group velocity tend to be lost. (b) Momentum distributions before (blue solid line) and after (orange dashed line) a loss event in the early time dynamics. (c) Difference of momentum distributions before and after the loss event. To obtain the momentum distributions, we set \(M=64\).~\label{fig:Zeno}}
\end{figure}

The emergence of the dip structure in the intermediate region can be understood as a quantum Zeno effect in the momentum space. 
At \(t=0\), there is no doubly-occupied site as depicted in Fig.~\ref{fig:Zeno} (a).
This means that in order for the loss events to happen, particles have to move with finite group velocity.
In other words, the loss event is more probable for faster particles.
Since the group velocity is the largest at \(|k|=\pi/2\) in the single-particle band of the one-dimensional Bose-Hubbard model, which is \(-2J\cos k\), the particles with \(|k|=\pi/2\) is the most likely to be lost.
In Figs.~\ref{fig:Zeno} (b) and (c), we compare the momentum distribution right before and after a loss event during early-time dynamics and see that the momentum distribution of the lost two particles is indeed peaked at \(|k| = \pi/2\).
As a consequence of series of such loss events, \(\braket{\hat{n}_k}/N\) forms the dips at \(|k| = \pi / 2\).
After the formation of the dip structure, the stronger dissipation for faster particles suppresses the redistribution of the particles towards states around \(|k| = \pi/2\).

\subsection{Strong suppression of particle transport}

The momentum distribution reflects the transition of the scaling law of the entanglement entropy, and is relatively easily accessible in ultracold-gas experiments.
However, the relation between the momentum distribution and the entanglement entropy seems unclear.
To resolve this difficulty, we propose a more direct signature of the entanglement transitions.

For this purpose, we borrow an idea from the experimental confirmations of the MBL states~\cite{schreiber_observation_2015,choi_exploring_2016}, which has utilized the breaking of ergodicity as an indicator of the area-law states.
In an area-law state, a part of a system does not possess an extensive entanglement entropy and thus cannot act as a thermal bath for the rest of the system~\cite{iyer_many-body_2013}.
The absence of the thermal bath results in the ergodicity breaking that is manifested, e.g., by the spatial imbalance of particles~\cite{schreiber_observation_2015,choi_exploring_2016}.
However, the true equilibrium state of this system is the vacuum state \(\ket{0}\) regardless of the scaling laws because the number of remaining particles continues to decrease.
Therefore, what one can observe is only dynamics towards a transient spatially imbalanced state. 
We expect that particle transport reflects a tendency toward the spatially imbalanced non-ergodic state.

In order to confirm this expectation, we simulate the following dynamics: In a 2\(M\)-site system, we prepare an initial state \(\ket{\psi_0} = \prod^M_{i=1} \hat{b}^\dagger_i\ket{0}\) in the left half of the system and set a high barrier potential \(100J\sum^{2M}_{i=M+1} \hat{n}_i\) which prevents the particles from coming into the right half.
We set \(U/J\) to \(5.0\).
After \(t=t_\mathrm{rel}\), we turn off the barrier potential and release the particles to the right half of the system. 
\(t_\mathrm{rel}\) is chosen to be within the time region where \(S_{\max}(t)\) takes almost a steady value.
We characterize particle transport by comparing the number of particles in the right half, i.e., \(N_\mathrm{r} = \sum^{2M}_{i=M+1}\braket{\hat{n}_i}\) with the half of the number of remaining particles \(N/2\).
Both \(N_\mathrm{r}\) and \(N/2\) can be measured in experiments~\cite{choi_exploring_2016}.

\begin{figure}[htp]
    \centering
    \includegraphics[width=\linewidth]{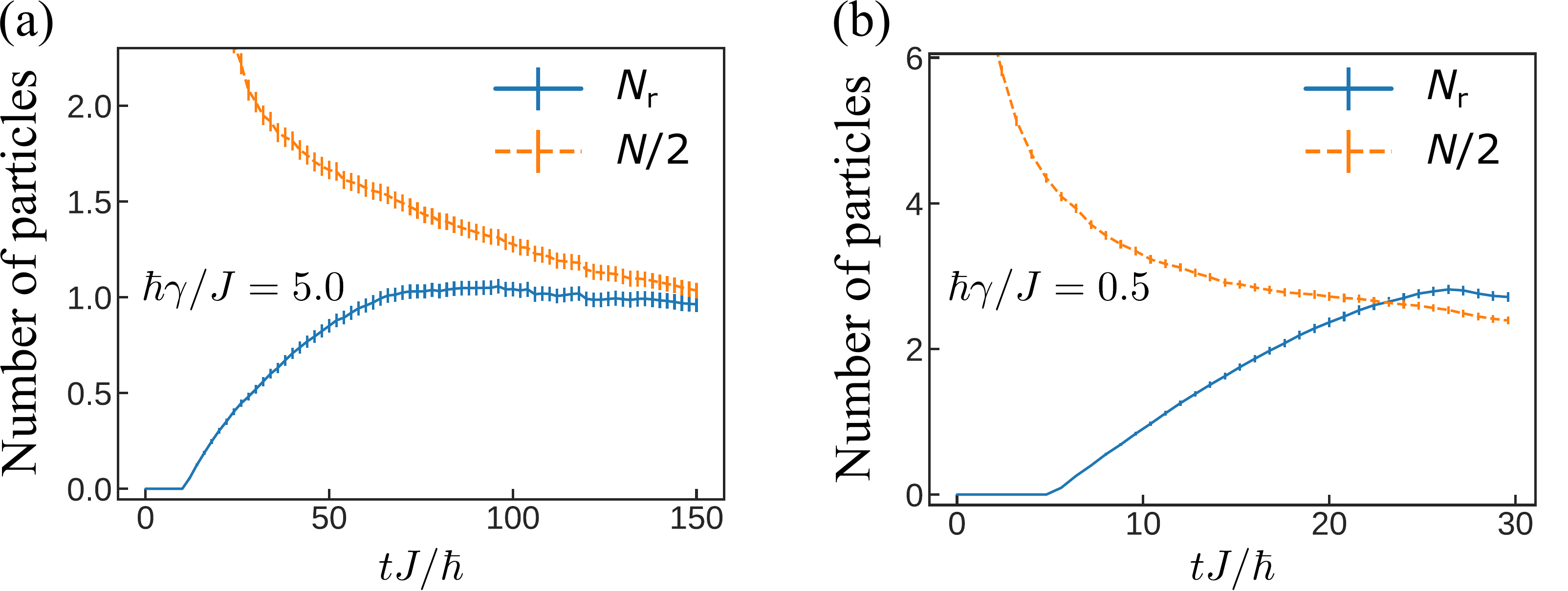}
    \caption{Time evolution of the number of particles in the right half of the system \(N_\mathrm{r}\) (blue solid line) and the half of the remaining number of particles  \(N/2\) (orange dashed line) for (a) \((\hbar \gamma / J, 2M, t_\mathrm{rel} J / \hbar) = (5.0, 128, 10)\) and (b) \((\hbar \gamma / J, 2M, t_\mathrm{rel} J / \hbar) = (0.5, 40, 5)\).\label{fig:barrier_release}}
\end{figure}

\begin{figure}
    \centering
    \includegraphics[width=\linewidth]{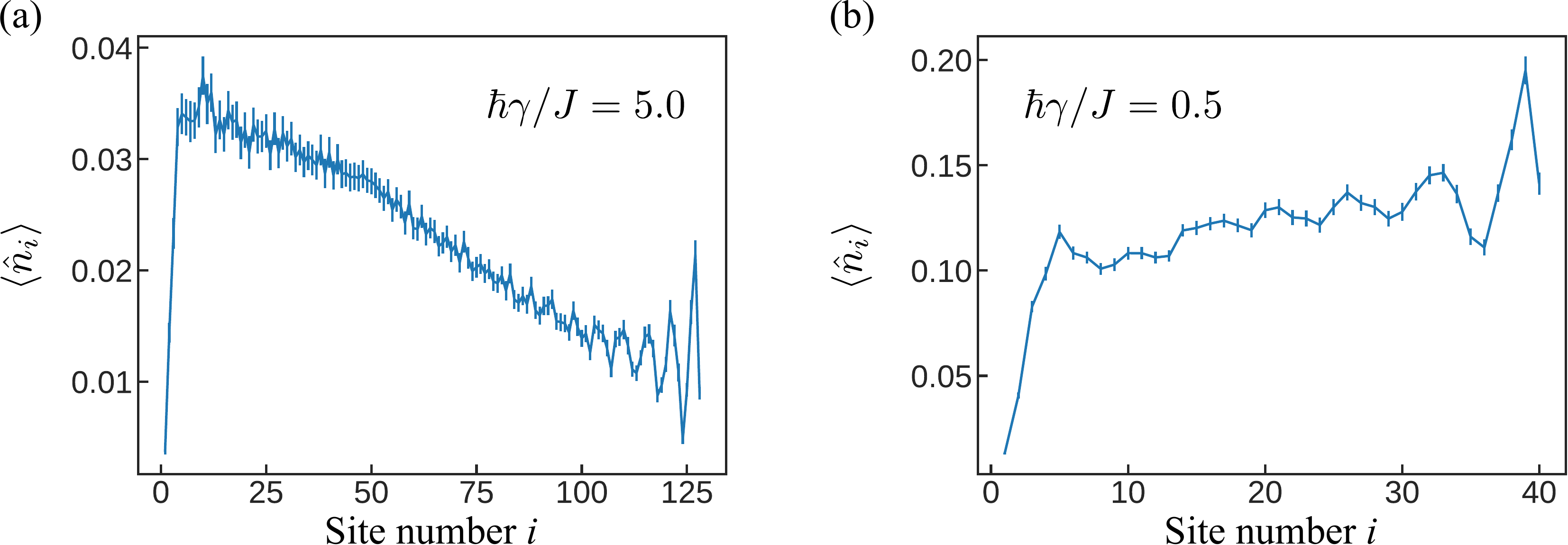}
    \caption{Spatial distribution of particles at \(t = t_\mathrm{rel}+\hbar M / J\) in the dynamics of (a) Fig.~\ref{fig:barrier_release}(a) and (b) Fig.~\ref{fig:barrier_release}(b).\label{fig:spatial_dis}}
\end{figure}

Figure~\ref{fig:barrier_release}(a) represents the time evolution of \(N_\mathrm{r}\) and \(N/2\) for \((\hbar \gamma / J, 2M, t_\mathrm{rel}J/\hbar) = (5.0, 128, 10)\), where the state at \(t=t_\mathrm{rel}\) is an ALSLC state.
For the broad region \(60 \lesssim t J / \hbar \lesssim 100\), there is a visible difference between \(N/2\) and a converged \(N_\mathrm{r}\).
The difference means that the delocalization of the particles is suppressed due to the dissipation, thus signaling the tendency toward a state without ergodicity.
By contrast, for \((\hbar \gamma / J, 2M, t_\mathrm{rel}J/\hbar) = (5.0, 128, 10)\), where the state at \(t=t_\mathrm{rel}\) is a volume-law state, \(N_\mathrm{r}\) exceeds \(N/2\) before the convergence as seen in Fig.~\ref{fig:barrier_release}(b). 
This overshoot behavior implies ballistic transport of the particles, which is consistent with the fact that in a volume-law state quantum information ballistically propagates.

The difference of tendencies is also visible as the inhomogeneity of the spatial distribution of particles, left-leaning or right-leaning, as shown in Fig.~\ref{fig:spatial_dis}.
In another volume-law region, say \(\hbar \gamma / J = 500.0\), the spatial distribution of particles and the time evolution of \(N_\mathrm{r}\) and \(N/2\) shown in Fig.~\ref{fig:gamma_500} are similar to those of \(\hbar \gamma / J = 0.5\) in terms of the ballistic propagation and the right-leaning spatial distribution of particles.
Thus, the strong suppression of particle transport clearly distinguishes the scaling law of the entanglement entropy as expected. 
For a fair comparison, we also present the dynamical behaviors of \(\hbar \gamma / J = 5.0\) case with system size \(2M=40\) in Fig.~\ref{fig:gamma_5}.
Even in the small system, the absence of the particle excess and the left-leaning spatial distribution of particles, which characterize the localization, are also visible.

\begin{figure}
    \centering
    \includegraphics[width=\linewidth]{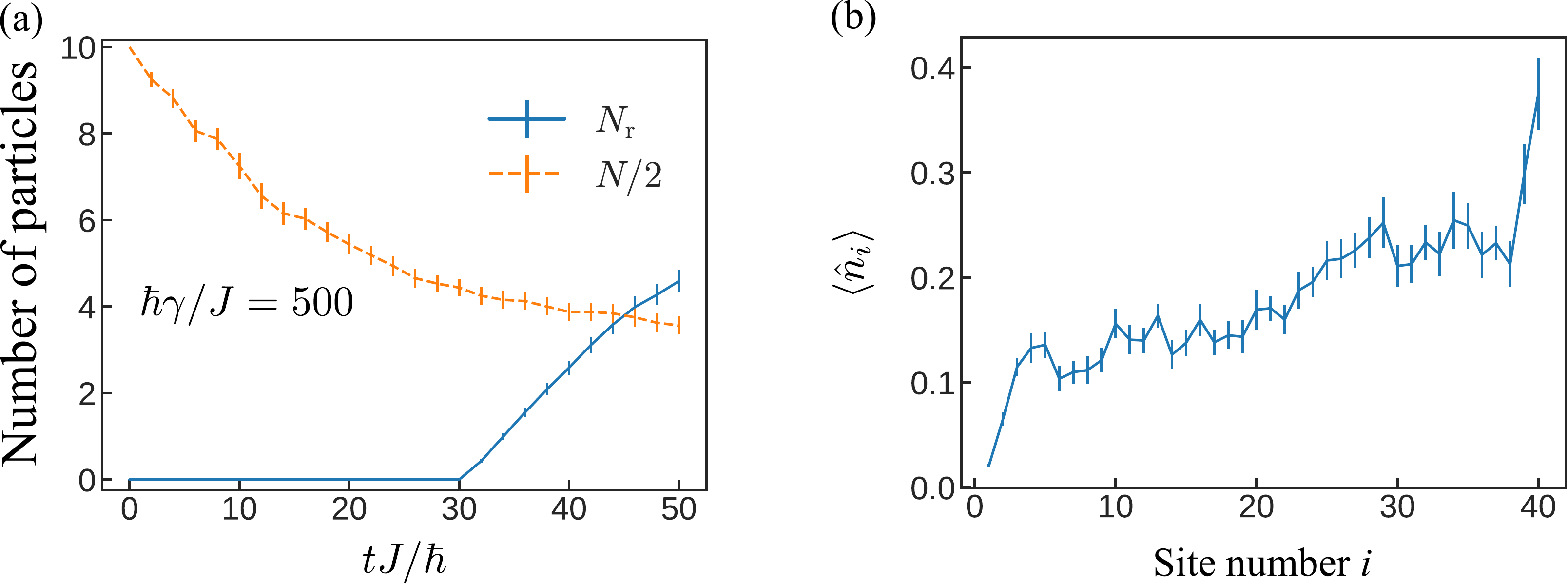}
    \caption{(a) Time evolution of the number of particles in the right half of the system \(N_\mathrm{r}\) (blue solid line) and the half of the remaining number of particles \(N/2\) (orange dashed line) for \((\hbar \gamma / J, 2M, t_\mathrm{rel}J/\hbar) = (500.0, 40, 30.0)\). (b) Spatial distribution of particles at  \(t = t_\mathrm{rel} + \hbar M / J\).\label{fig:gamma_500}}
\end{figure}

\begin{figure}
    \centering
    \includegraphics[width=\linewidth]{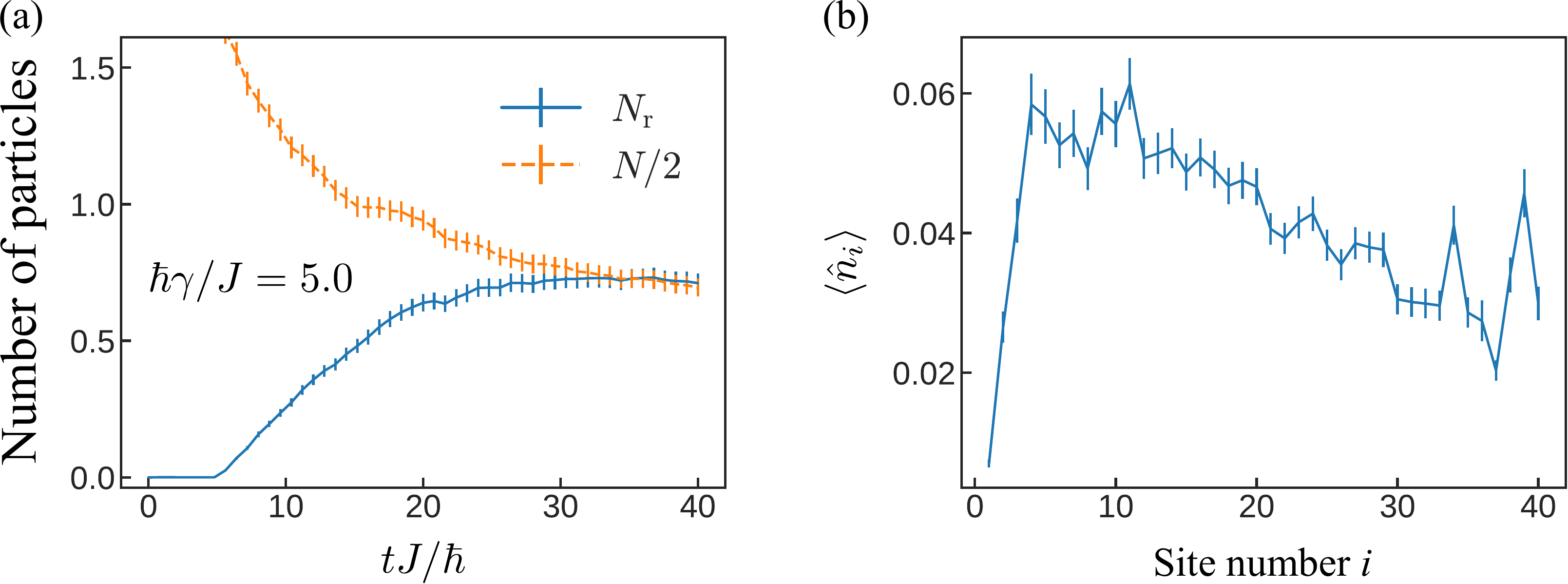}
    \caption{(a) Time evolution of the number of particles in the right half of the system \(N_\mathrm{r}\) (blue solid line) and the half of the remaining number of particles \(N/2\) (orange dashed line) for \((\hbar \gamma / J, 2M, t_\mathrm{rel}J/\hbar) = (5.0, 40, 5.0)\). (b) Spatial distribution of particles at  \(t = t_\mathrm{rel} + \hbar M / J\).\label{fig:gamma_5}}
\end{figure}

\begin{figure}
    \centering
    \includegraphics[width=0.8\linewidth]{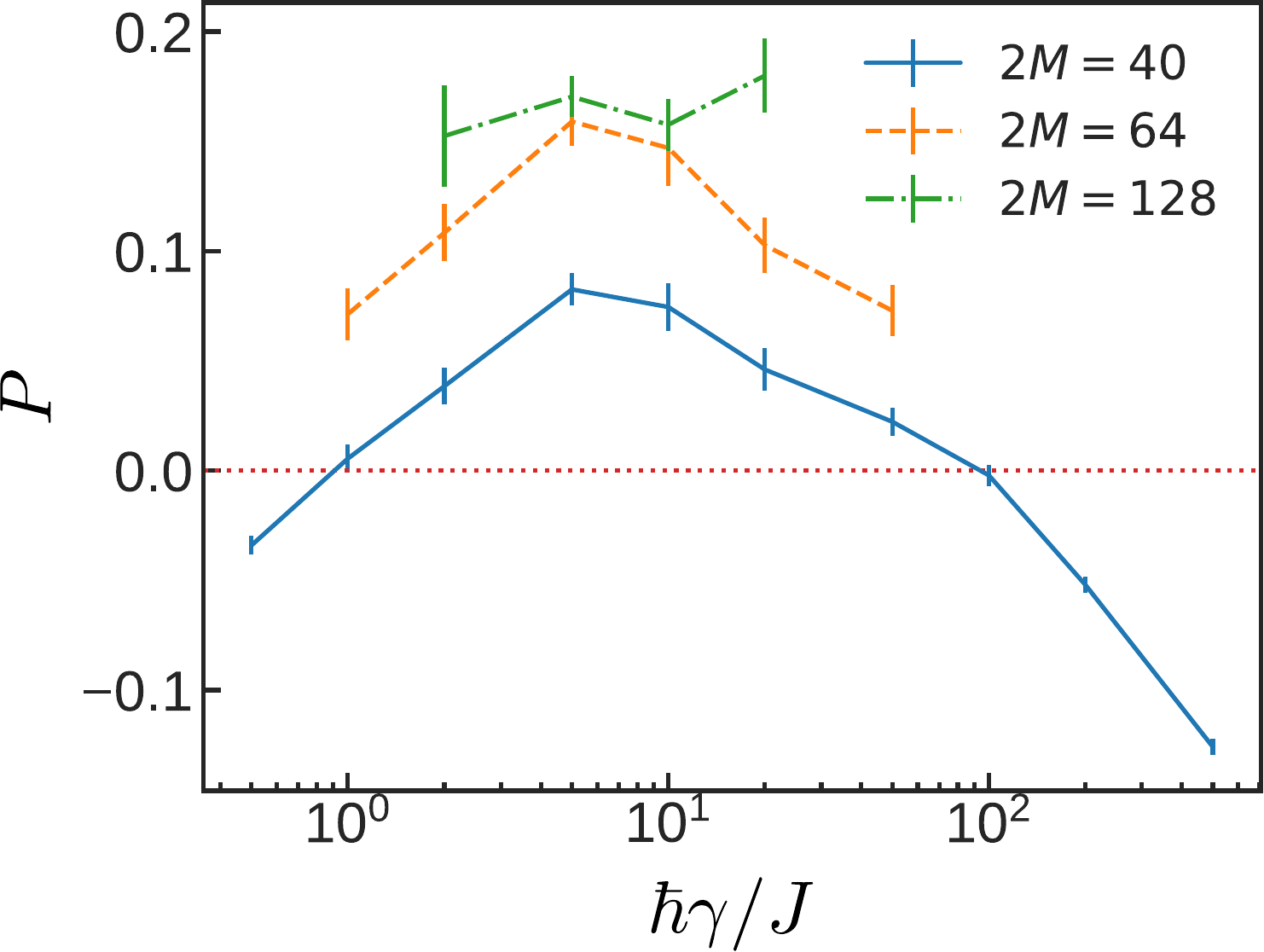}
    \caption{The imbalance \(P\) versus the dissipation strength \(\hbar \gamma / J\) at \(t_\mathrm{obs} = t_\mathrm{rel} + \hbar M / J\).
    For the simulation with \(2M = 40\), we use \(t_\mathrm{rel}J/\hbar = 5.0\) for \(\hbar \gamma / J \leq 5.0\) and set \(t_\mathrm{rel}J/\hbar \) to 6.0, 10.0, 15.0, 20.0, and 30.0 for \(\hbar \gamma / J = 10.0, 20.0, 50.0, 100.0\), and 500.0, respectively. For \(2M=64\) (128) case, we use \(t_\mathrm{rel}J/\hbar = 10.0\) for \(\hbar \gamma / J \leq 20.0\) (10.0) and \(t_\mathrm{rel}J/\hbar = 20.0\) for \(\hbar \gamma / J = 50.0\) (20.0).~\label{fig:imbalance}}
\end{figure}
In order to quantify how much particle transport is suppressed, we calculate the imbalance between \(N_\mathrm{r}\) and \(N/2\) defined as 
\begin{align}
    P = \frac{N/2 - N_\mathrm{r}}{N/2 + N_\mathrm{r}}
\end{align}
at \(t_\mathrm{obs} = t_\mathrm{rel} + \hbar M / J\).
\(t_\mathrm{obs} - t_\mathrm{rel}\) corresponds to a rough estimate of the time scale in which the particles released at \(t=t_\mathrm{rel}\) reaches the right edge of the system.
If \(P\) significantly exceeds zero, the state at \(t=t_\mathrm{rel}\) is an ALSLC state.
If \(P \leq 0\), that is a volume-law state.
Otherwise, the state lies in an intermediate regime.
Figure~\ref{fig:imbalance} represents \(P\) versus \(\hbar \gamma / J\) for \(2M = 40\), 64, and 128.
The distinction between the volume-law and ALSLC states made from \(P\) is consistent with the scaling law shown in Fig.~\ref{fig:Size_dep}.
We also see that the imbalance becomes more visible as the system size increases.
Although it is quite difficult to access the volume-law region in the case of the larger systems (\(2M=64\) and 128) with numerical simulations, it is expected that \(P \leq 0\) with the ballistic particle transport regardless of the system size.
In short, the imbalance \(P\) serves as an indicator of whether the initial state of the release dynamics is the ALSLC or the volume-law state, which can be observed in experiments.

\section{Summary\label{sec:summary}}
We proposed the measurement-induced transitions (MITs), which have been theoretically found in recent studies of quantum circuit models~\cite{li_quantum_2018,chan_unitary-projective_2019,skinner_measurement-induced_2019,szyniszewski_entanglement_2019,jian_measurement-induced_2020,gullans_dynamical_2019,li_measurement-driven_2019,choi_quantum_2020,bao_theory_2020,tang_measurement-induced_2020}, can be experimentally observed by using ultracold bosons in optical lattices with controllable dissipation.
We employed a quasi-exact numerical method to investigate effects of dissipation on quench dynamics of the one-dimensional Bose-Hubbard model with a two-body loss term.
By computing the maximal entanglement entropy of the system during the time evolution, we found two MITs.
Specifically, when the strength of the dissipation increases, the scaling of the entanglement changes from a volume law to an area law with a logarithmic correction, and again to the volume law.
We showed that the momentum distribution, a standard observable in ultracold-gas experiments, reflects the change of the scaling laws.
We also suggested that the strong suppression of particle transport in the dynamics after release of the particles to an empty space are more direct observable signatures in experiments for distinguishing the area-law states from the volume-law states.

We could not locate precisely the critical points for the two MITs because it was impossible to efficiently describe the volume-law states of the dissipative Bose-Hubbard model with currently available numerical techniques.
Since in experiments with ultracold gases the tractable system size is not limited by the volume-law entanglement, the determination of the critical points will be a meaningful target of quantum simulations.

\begin{acknowledgments}
We thank S. Nakajima, Y. Takahashi, Y. Takasu, and T. Tomita for fruitful discussions.
The MPS calculations in this work are performed with ITensor library, http://itensor.org.
This work was financially supported by KAKENHI from Japan Society for the Promotion of Science: Grant No. 18K03492 and No. 18H05228, by  CREST, JST No. JPMJCR1673, and by MEXT Q-LEAP Grant No.\@ JPMXS0118069021.
\end{acknowledgments}
 
\end{document}